\documentclass{PoS}

\usepackage{amsmath}
\usepackage{amssymb}
\usepackage{amsfonts}
\makeatletter
    \newcommand{\msbar}{\overline{\mathrm{MS}}}
    \newcommand{\Tabref}[1]{Table~\ref{#1}}
    \newcommand{\Eqref}[1]{Eq.~(\ref{#1})}
    \newcommand{\Figref}[1]{Fig.~\ref{#1}}
  \makeatother

  \title{
  \begin{picture}(0,0)(0,0)
  \put(350,75){\makebox(0,0)[l]{\textnormal{\normalsize KEK-CP-373}}}
  \end{picture}
  Renormalization of bilinear and four-fermion operators through temporal moments}

  \ShortTitle{Renormalization of bilinear and four-fermion operators through temporal moments}

  \author{\speaker{Tsutomu Ishikawa}$^{a,b}$,\ Katsumasa Nakayama$^{b,c}$,\ Shoji Hashimoto$^{a,b}$ \\
          $^{a}$Graduate University for Advanced Studies (SOKENDAI), Tsukuba 305-0801, Japan\\
          $^{b}$KEK Theory Center, Institute of Particle and Nuclear Studies, High Energy Accelerator Research Organization (KEK), Tsukuba 305-0801, Japan\\
          $^{c}$NIC, DESY Zeuthen, Platanenallee 6, 15738 Zeuthen, Germany\\
          E-mail: \email{tsuto@post.kek.jp}
          }

  \abstract{We propose a renormalization scheme that can be simply implemented on the lattice. It consists of the temporal moments of two-point and three-point functions calculated with finite valence quark mass. The scheme is confirmed to yield a consistent result with another renormalization scheme in the continuum limit for the bilinear operators. We apply a similar renormalization scheme for the non-perturbative renormalization of four-fermion operators appearing in the weak effective Hamiltonian.}

  \FullConference{37th International Symposium on Lattice Field Theory - Lattice2019\\
          16-22 June 2019\\
          Wuhan, China}

\begin{document}

  \section{Introduction}
   $B\rightarrow K^{(*)} l^+ l^- $ decay is one of the flavor-changing neutral current processes. Its decay amplitude in the Standard Model is suppressed by the GIM mechanism, and is sensitive to new physics. In the theoretical analysis, however, charmonium long-distance effects make it difficult to accurately predict the Standard Model contributions.
   Lattice QCD may be able to treat such effects from the first-principles (for instance, a test of factorization is attempted in \cite{Nakayama:2019eth}).\par
   In the study of weak decays on the lattice, renormalization is necessary. Since most of the lattice operators have logarithmic or power divergences toward the continuum limit, we should remove the divergences and give the proper scale dependences. Even when the operator does not have a divergence such as the case of vector current, we need to take the discretized effect into account. We can obtain the correct physical quantity only after the renormalization.
  \par
  Renormalization can be performed by applying a matching directly or indirectly. For the quantities to be matched we require the following properties: typical length scale is short enough to use perturbation theory. At the same time the quantity has to allow precise lattice calculation. Then, we match the lattice calculation with the corresponding perturbative (usually in the $\mathrm{\overline{MS}}$ scheme) calculation. For instance, the coordinate space correlators $ G(x)= \langle 0|T J(x)J(0)| 0 \rangle$ at a short (but nonzero) distance $x$
   are used in the X-space method.
  Another example is the RI/MOM scheme, where the vertices with external free quark lines are matched to the corresponding $\msbar$ calculations.
   \par
   In the present work, we propose a new matching procedure to determine the renormalization factors for lattice operators. It is based on the temporal moments of charmonium correlators. We match a temporal moment with the $\msbar$ counterpart or with tree level amplitude.  A similar method for the vector current has been studied in the literature \cite{Donald:2012ga, Colquhoun:2014ica}. We apply the method for pseudoscalar operators and verify that our method works well. Then we extend the method to four-fermion operators to describe weak decays.

   \section{Renormalization of bilinear operators}
   In the continuum theory, moments $M_k$ of a charmonium correlator
   are defined by a $q^2$ derivative of the vacuum polarization function $\Pi (q^2) $ at $q^2=0$:
  \begin{align}
    q^2 \Pi (q^2) &= \int d^4x \ e^{iqx} \langle 0|T j_5(x)j_5(0)| 0 \rangle,\\
    M_k&=\left.\frac{\partial^k \Pi(q^2)}{\partial (q^2)^k}\right|_{q^2=0},
  \end{align}
  where $j_5(x)=i \bar{\psi}_{c}(x) \gamma_{5} \psi_{c}(x)$ is the charmonium pseudoscalar density operator. From dimensional analysis, the moments do not contain
  any extra divergence due to $x \rightarrow 0$ for $k>1$. We use these finite quantities to determine the renormalization factor $Z^{\msbar/\mathrm{lat}}(\mu,a) $ for the lattice operator, i.e. we impose a  matching condition for the moments at a renormalization scale $\mu$:
  \begin{align}
    \label{eq:def_renorm_cond}
    \left.\frac{\partial^k }{\partial (q^2)^k}\Pi^{\msbar}(\mu;q^2)\right|_{q^2=0}=\left(Z^{\msbar/\mathrm{lat}}(\mu,a)\right)^2\left.\frac{\partial^k }{\partial (q^2)^k}\Pi^{\mathrm{lat}}(a;q^2)\right|_{q^2=0},
  \end{align}
  where $\Pi^{\msbar}(\mu;q^2)$ and $\Pi^{\mathrm{lat}}(a;q^2)$ are vacuum polarization functions in the $\msbar$ scheme and on the lattice, respectively.
  The perturbative expansion of the moments are known to $\mathcal{O}(\alpha_s^3)$ in the $\msbar$ scheme  \cite{Maier:2009fz}.\par
   The $q^2$-derivative on the lattice that appears on r.h.s. of \Eqref{eq:def_renorm_cond}  is equivalent to a temporal moment of the charmonium correlation function.
   The charmonium correlation function is written on the lattice as
   \begin{align}
     G(t)&=a^6\sum_{\vec{x}} \langle j_5(t,\vec{x})j_5(0,0) \rangle .
   \end{align}
   Then, the temporal moment of the correlator is given as
   \begin{align}
     G_n  &= \sum_t \left( \frac{t}{a} \right) ^n G(t)=\left.\frac{\partial^k }{\partial (q^2)^k}\Pi^{\mathrm{lat}}(a;q^2)\right|_{q^2=0},
   \end{align}
   where $n$ is related to $k$ in (\ref{eq:def_renorm_cond}) as $n=2k+2$.
   On the lattice, the time $ t/a$ runs from $ -T/2a + 1$ to $ T/2a$, and correlators are even functions of  time. Typical length scale is given by an inverse of the charm quark mass $m_c^{-1}$, which is short enough to describe perturbatively. The temporal moment on the lattice has been shown to provide  precise determination of charm quark mass and strong coupling constant \cite{Allison:2008xk,McNeile:2010ji,Nakayama:2016atf}, which indicate that it may be used for the purpose of renormalization \cite{Donald:2012ga, Colquhoun:2014ica}. In practice, we divide the moments by their one-loop calculation (or the vacuum polarization function at $\mathcal{O}(\alpha_s^0)$) and multiply the charmonium mass to reduce discretized errors as in \cite{Allison:2008xk,McNeile:2010ji,Nakayama:2016atf}.

   We use ensembles with $\mathrm{N_f=2+1} $
    M\"obius domain-wall fermions.
    The parameters of our lattice simulations are shown in \Tabref{tb:ensemble}.
     We also input the strong coupling constant
     $\alpha _s(2\ \mathrm{GeV} )=0.3022$  and
      charm quark mass $\bar{m}_c(2\ \mathrm{GeV})= 1.09\ \mathrm{GeV}$, which are obtained by solving renormalization group equations from the PDG average $\alpha _s(M_z )=0.1181$ and $\bar{m}_c(\bar{m}_c)\ $ $=$
        $1.27 \ \mathrm{GeV}$ \cite{Tanabashi:2018oca}.\par
   \begin{table}[t]
     %\vspace{10pt}
     \centering
     \begin{tabular}{ccc|ccc|c}
       $\beta$ &   $ a^{-1}\ [\mathrm{GeV}]$  & $L^3\times T (\times L_5 )$ & \#meas &$am_{ud}$&$am_s$&$am_c$\\ \hline
       4.17& 2.453(4)&$32^3 \times 64 (\times 12) $ &100& 0.007&0.040&0.44037\\
       4.35& 3.610(9)&$48^3 \times 96 (\times 8) $ &50& 0.0042&0.0250&0.27287\\
       4.47& 4.496(9)&$64^3 \times 128 (\times 8) $ &50& 0.0030&0.015&0.210476\\
   \end{tabular}
   \caption{Ensembles for our simulations. $m_{ud}$ and $m_s$ are masses of sea quarks and $m_c$ is a valence quark mass for each ensemble.}
     \label{tb:ensemble}
   \end{table}

\begin{figure}[t]
   \begin{center}
     \includegraphics[width=10cm ]{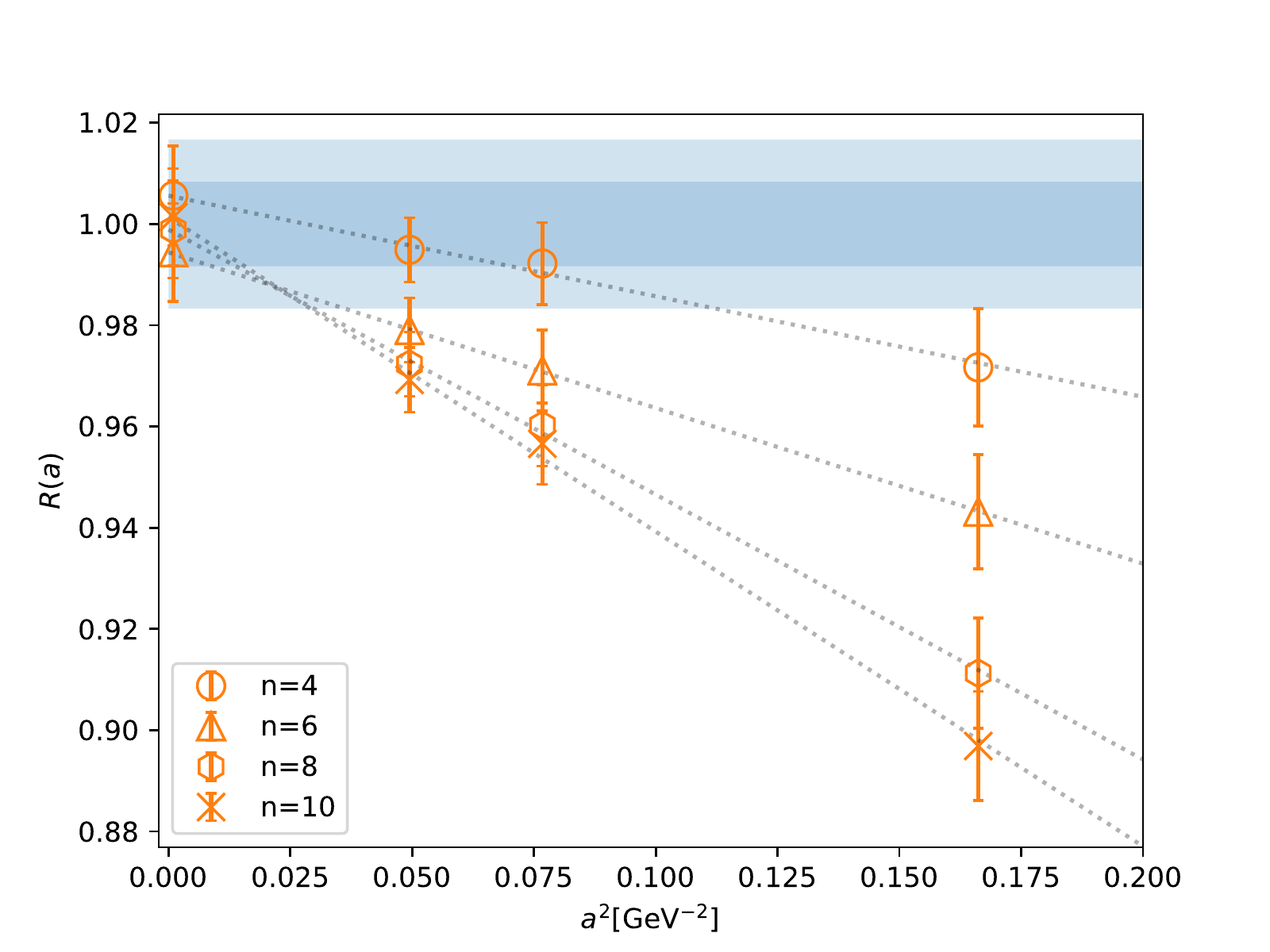}
     \caption{Ratio of renormalization contstants for the pseudoscalar density operator with our method to ones with X-space method defined in \Eqref{eq:ratio_xspace}. The thick band represents  an estimate of unknown $\mathcal{O}(\alpha_s^4)$ correction given by $1 \pm \alpha_s^4$ and the thin band shows $1 \pm 2\alpha_s^4$.}
     \label{fig:ratio_mywork_tomii}

   \end{center}
   \end{figure}

   We check the consistency with another method, taking the renormalization constant from the X-space method as a reference \cite{Tomii:2016xiv}. We calculate the ratio of our results at a renormalization scale $\mu =2\ \mathrm{GeV}$ to the reference:
  \begin{align}
    \label{eq:ratio_xspace}
    R(a)=Z_P^{(\mathrm{moment})}(\mu=2\ \mathrm{GeV},a)/Z_P^{(\mathrm{X \mathchar`- space})}(\mu=2\ \mathrm{GeV},a),
  \end{align}
  where $Z_P^{(\mathrm{moment})}$ is the renomalization constant calculated by our method and $Z_P^{(\mathrm{X \mathchar`- space})}$ is the renormalization constant obtained by the X-space method. \Figref{fig:ratio_mywork_tomii} shows the ratio. After taking the continuum limit, they are consistent with each other, i.e. $R(a)=1$, up to truncation errors of $\mathcal{O}(\alpha_s^4)$ and discretization error of $\mathcal{O}(a^2)$.
    \par
\begin{figure}[t]
    \begin{center}
      \begin{tabular}{c}      % 2
        \begin{minipage}{0.47\hsize}
          \begin{center}
            \includegraphics[ width=7cm]{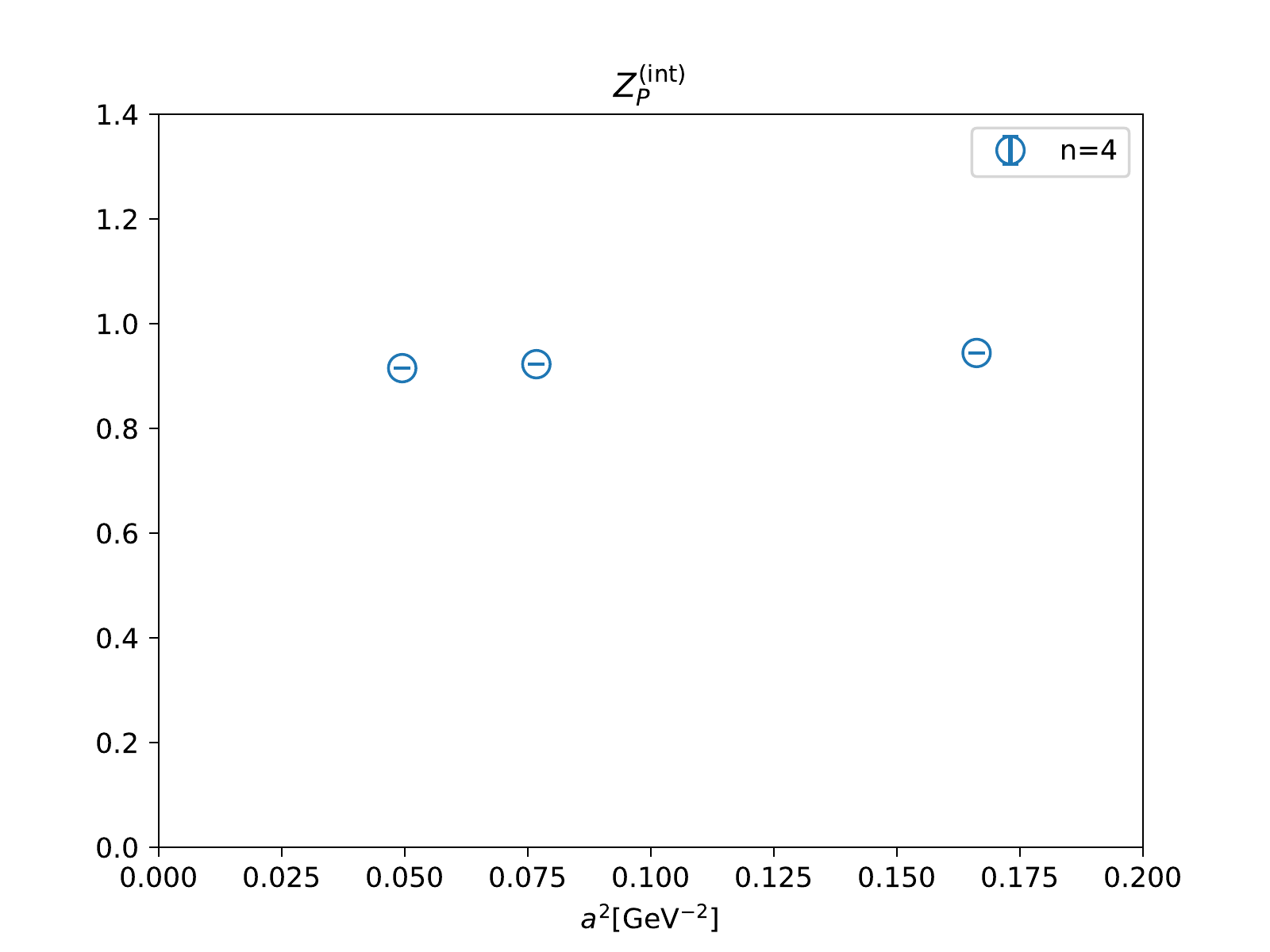}
            \hspace{1.6cm}
          \end{center}
        \end{minipage}

        \begin{minipage}{0.47\hsize}
          \begin{center}
            \includegraphics[width=7cm]{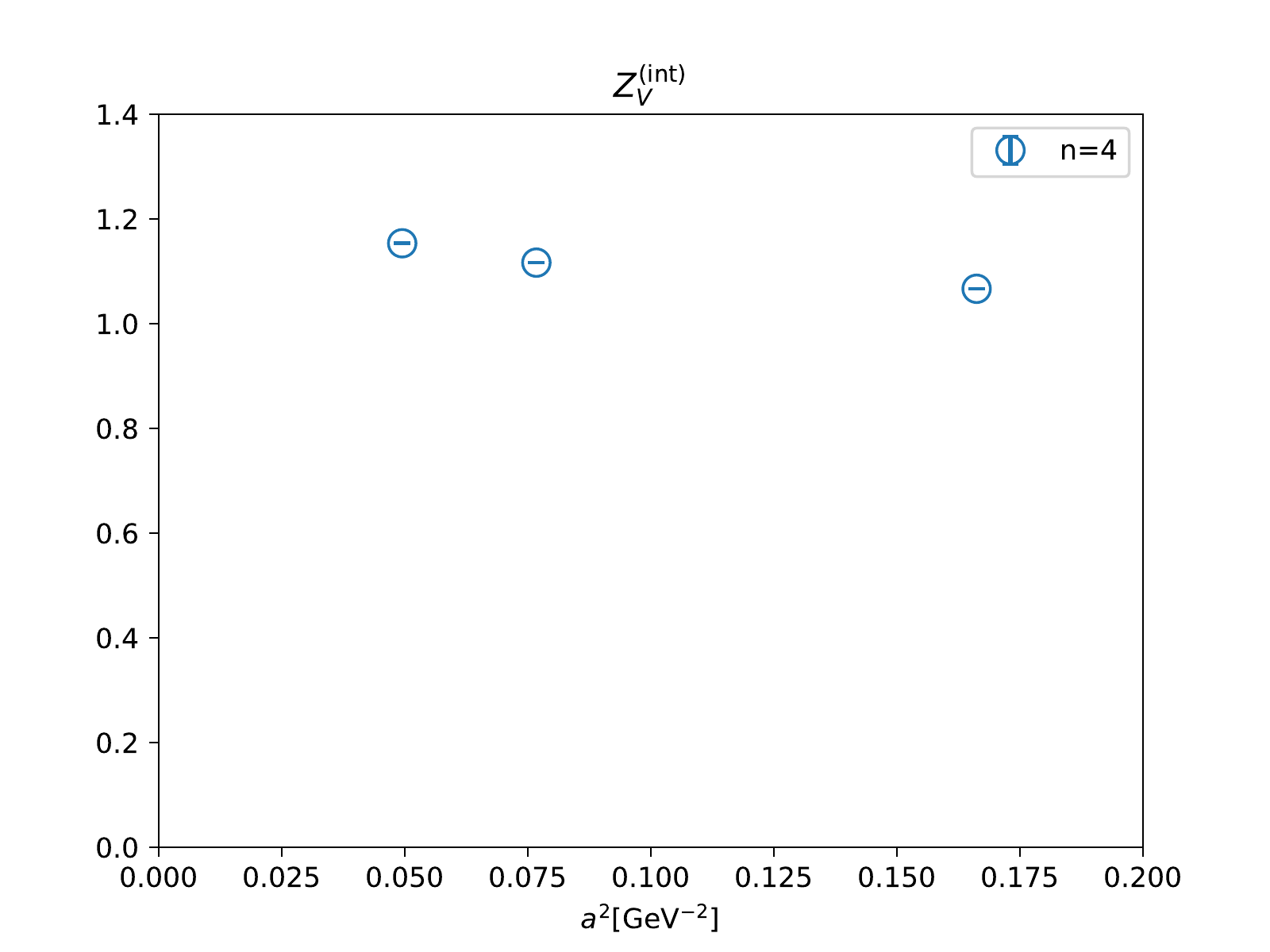}
            \hspace{1.6cm}
          \end{center}
        \end{minipage}
      \end{tabular}
      \small
      \caption{Renormalization constants in the intermediate scheme. Left and Right show the constants for pseudoscalar and vector channels, respectively.}
      \label{fig:renorm_int}

    \end{center}
    \end{figure}

    Temporal moments can also be used as an  intermediate renormalization scheme.
    The renormalization constant $Z^{(\mathrm{int})}$ is defined through a matching with the moments at the tree level or $\mathcal{O}(\alpha_s^0)$:
    \begin{align}
      \label{eq:renorm_int}
      \left.\frac{\partial^k }{\partial (q^2)^k}\Pi^{\mathrm{tree}}(q^2)\right|_{q^2=0}=\left(Z^{(\mathrm{int})}(a)\right)^2\left.\frac{\partial^k }{\partial (q^2)^k}\Pi^{\mathrm{lat}}(a;q^2)\right|_{q^2=0}.
    \end{align}
   The results are shown in \Figref{fig:renorm_int}. We show only the case of $n=4$ (or $k=1$) since it is  the lowest order moment, which is more dominated by short-distance physics and suitable for renormalization. This scheme is applicable independently of the channel.

   \section{Extension to four-fermion operators}
    We extend our method to four-fermion operators, appearing in the effective weak Hamiltonian. In particular, we focus on the operators $O_1$ and $O_2$ that represent the charmonium contribution in $B\rightarrow K^{(*)} \ l^+l^-$ decays:
    \begin{align}
      \mathcal{H}_{\mathrm{eff}}&=\frac{G_{F}}{\sqrt{2}} V_{c s}^{*} V_{c b}\left(C_{1} O_{1}+C_{2} O_{2}\right),\\
       O_{1} &=\left(\overline{s}_{i} \gamma_{\mu} P_{-} c_{j}\right)\left(\overline{c}_{j} \gamma_{\mu} P_{-} b_{i}\right), \\ O_{2} &=\left(\overline{s}_{i} \gamma_{\mu} P_{-} c_{i}\right)\left(\overline{c}_{j} \gamma_{\mu} P_{-} b_{j}\right),
    \end{align}
    where $G_F,\ V_{cs,cb},\ C_{1,2}$ and $P_{-}=(1-\gamma_5)/2$ are the Fermi constant, elements of the CKM matrix, a Wilson coefficient of $O_{1,2}$ and a left-handed projection operator, respectively. $i$ and $j$ are color indices.
    \par
\begin{figure}[t]
  \begin{center}
    \includegraphics[width=12cm ]{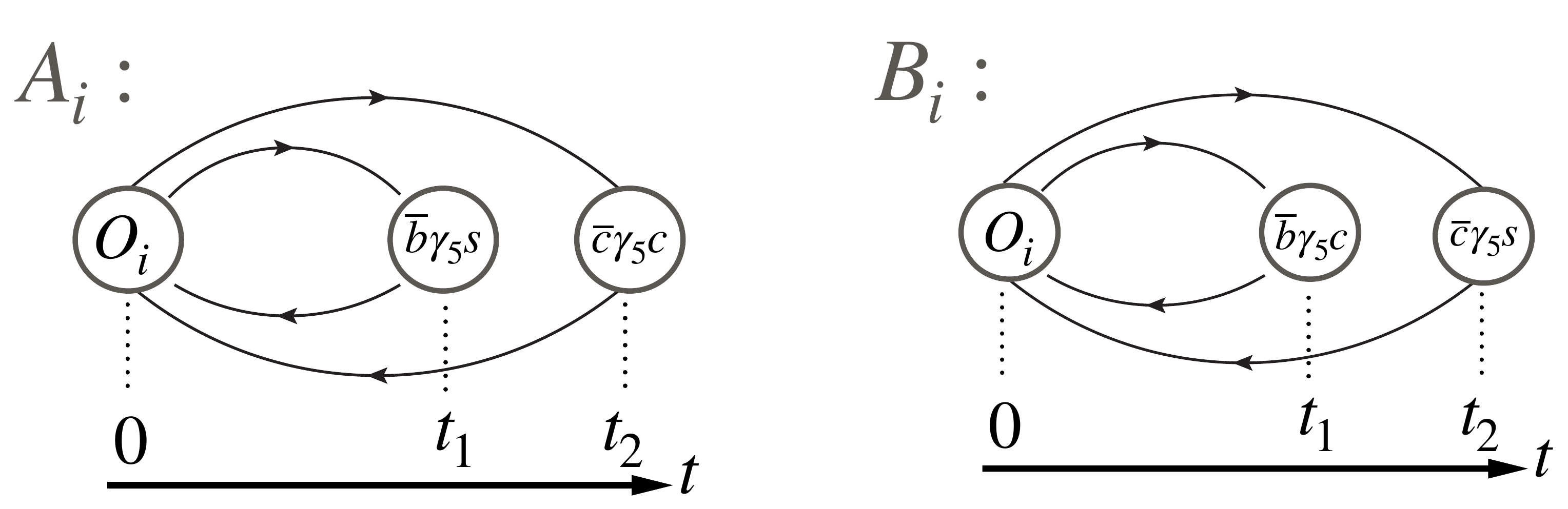}
    \caption{Two correlators to calculate temporal moments.}
    \label{fig:correlators}

  \end{center}
  \end{figure}

   The operators $O_{1,2}$ mix through the renormalization, and the renormalization constants form a $2\times 2$ matrix. The relation between the renormalized operators $O_i^R$ and the bare lattice operators $O_i$ is
   \begin{align}
     \begin{pmatrix}
       O^R_1 \\ O^R_2
     \end{pmatrix}=
     \begin{pmatrix}
       Z_{11}& Z_{12}\\Z_{21}& Z_{22}
     \end{pmatrix}
     \begin{pmatrix}
       O_1\\O_2
     \end{pmatrix}.
   \end{align}
   Here we do not have to consider a mixing with lower dimensional operators since they do not create  $ c \bar{c} $ states without involving disconnected diagrams which we neglect in this work.
  To determine the renormalization constants, we prepare two external states for each operator and calculate correlation functions with them placed at $t_1$ and $t_2$ as depicted in \Figref{fig:correlators}:
  \begin{align}
    A_i(t_1,t_2)&=a^{12}\sum_{\vec{x_1},\vec{x_2}}\left\langle\left(\bar{b}\gamma_5s\right)(t_1,\vec{x}_1) O_i(0,\vec{0})\left(\bar{c}\gamma_5c\right)(t_2,\vec{x}_2)\right\rangle,\\
    B_i(t_1,t_2)&=a^{12}\sum_{\vec{x_1},\vec{x_2}}\left\langle\left(\bar{b}\gamma_5c\right)(t_1,\vec{x}_1) O_i(0,\vec{0})\left(\bar{c}\gamma_5s\right)(t_2,\vec{x}_2)\right\rangle.
  \end{align}

  The renormalization constants are obtained by matching the double temporal moments of the form
  \begin{align}
    A_i^{(n_1,n_2)}&=(am_c)^2\sum_{t_1,t_2}\left(\frac{t_1}{a}\right)^{n_1}\left(\frac{t_2}{a}\right)^{n_2}A_i(t_1,t_2), \\
    B_i^{(n_1,n_2)}&=(am_c)^2\sum_{t_1,t_2}\left(\frac{t_1}{a}\right)^{n_1}\left(\frac{t_2}{a}\right)^{n_2}B_i(t_1,t_2).
  \end{align}
  The factor $(am_c)^2$ cancels the renormalization factor of the pseudoscalar density operator introduced  for the external source.
  The orders of the moments $n_1$ and $n_2$ must be odd, otherwise the moments vanish at the tree level. To avoid any extra divergence, $n_1$ and $n_2$ must be larger than 2. We therefore take $n_1 = n_2 =3$, which provides  the shortest distance correlation.\par

  We impose a renormalization condition on the moments:
  \begin{align}
    \label{eq:renorm_a}
  \left.A_i^{(3,3)}\right|_{\mathrm{renorm.}}&=\left.A_i^{(3,3)}\right|_{\mathrm{tree}},\\
  \label{eq:renorm_b}
  \left.B_i^{(3,3)}\right|_{\mathrm{renorm.}}&=\left.B_i^{(3,3)}\right|_{\mathrm{tree}},
  \end{align}
  where the l.h.s. is the renormalized quantities and the r.h.s. is the moments at the tree level.  We set all valence quark masses $m_b= m_c= m_s$ to $m_c$ because the renormalization constants should be determined by the UV behavior and can be made independent of each quark mass.
  The bulk of those moments is given in the short-distance regime, and we control the distance scale by setting the heavy valence quark mass.
  We can simplify (\ref{eq:renorm_a}) and (\ref{eq:renorm_b}) using Fierz identities for $ O_1$. Namely,the Fierz transformations of $O_1$ gives
  \begin{align}
    &\left\langle\left(\overline{b}\gamma_5s\right)\left(\overline{s}_{i} \gamma_{\mu} P_{-} c_{j}\right)\left(\overline{c}_{j} \gamma_{\mu} P_{-} b_{i}\right)\left(\overline{c}\gamma_5c\right)\right\rangle
    =\left\langle\left(\overline{b}\gamma_5s\right)\left(\overline{s}_{i} \gamma_{\mu} P_{-} b_{i}\right)\left(\overline{c}_{j} \gamma_{\mu} P_{-} c_{j}\right)\left(\overline{c}\gamma_5c\right)\right\rangle,
  \end{align}
  which is equal to
  \begin{align}
  \left\langle\left(\overline{b}\gamma_5c\right)O_2\left(\overline{c}\gamma_5s\right)\right\rangle
  =\left\langle\left(\overline{b}\gamma_5c\right)\left(\overline{c}_{j} \gamma_{\mu} P_{-} b_{j}\right)\left(\overline{s}_{i} \gamma_{\mu} P_{-} c_{i}\right)\left(\overline{c}\gamma_5s\right)\right\rangle
  \label{eq:B_2_Fierz}
  \end{align}
  when the valence quark masses are degenerate.
   Note that we neglect the disconnected diagrams.
   We then obtain two identities:
   \begin{align}
     A_1^{(n_1,n_2)}=B_2^{(n_1,n_2)},\
     A_2^{(n_1,n_2)}=B_1^{(n_1,n_2)},
   \end{align}
    and $  Z_{11}=Z_{22},\  Z_{12}=Z_{21}.$
    As a consequence, it is sufficient to solve a linear equation
    \begin{align}
      Z_{11}A_1^{(3,3)}+Z_{12}A_2^{(3,3)}&=\left.A_1^{(3,3)}\right|_{\mathrm{tree}},\\
      Z_{11}A_2^{(3,3)}+Z_{12}A_1^{(3,3)}&=\left.A_2^{(3,3)}\right|_{\mathrm{tree}},
    \end{align}
    with $A_1^{(3,3)}$ and $A_2^{(3,3)}$ as inputs.
    \par
    The numerical results are shown in \Tabref{tb:result_ff}.
    The anomalous dimension can also be calculated by taking a difference between two nearby lattice spacings
    \begin{align}
      \gamma_{ij}&=-a\frac{\partial}{\partial a}\log{Z}_{ij}
  =-Z^{-1}_{ik}a\frac{\partial}{\partial a}Z_{kj}.
    \end{align}
    We find that the signs are consistent with one-loop results.
    \begin{table}[t]
      %\vspace{10pt}
      \centering
      \begin{tabular}{c|c|c|c}
        $\beta$ &   $ a^{-1}\ [\mathrm{GeV}]$  & $Z_{11}$ &$Z_{12}$\\ \hline
        4.17& 2.453(4)&0.754(8)  &0.072(2)\\
        4.35& 3.610(9)&0.669(11) &0.093(4)\\
        4.47& 4.496(9)&0.645(15) &0.098(4)\\
    \end{tabular}
    \caption{Results from our method. The errors of renormalization constants include only statistical ones.}
      \label{tb:result_ff}
    \end{table}
  \section{Discussions}
  We propose a renormalization scheme based on the charmonium moments. Our method requires no gauge fixing, unlike the RI/MOM scheme. We confirm that the scheme yields a consistent result with the X-space scheme up to truncation and discretization errors for a pseudoscalar density operator. We extend the method to the four-fermion operators.

  \section*{Acknowledgement}
  Numerical computations are performed on Oakforest-PACS at JCAHPC.
This work was supported in part by JSPS KAKENHI Grant Number JP18H03710
and by MEXT as "Priority Issue on post-K computer".
K. N. is supported by the Grant-in-Aid for JSPS (Japan Society for the Promotion of Science) Research Fellow (No. 18J11457).
  \bibliography{lattice2019}
  \bibliographystyle{unsrt}

  \end{document}